\documentclass[twocolumn,preprintnumbers,amsmath,amssymb]{revtex4}

\usepackage{graphicx}
\usepackage{dcolumn}
\usepackage{bm}
\newcommand{\theq}{\theta_{\text{eq}}}
\newcommand{\vct}[1]{\mathbf{#1}}

\begin{document}

\title{Motion of nanodroplets near chemical heterogeneities }
\author{A.  Moosavi} 
\affiliation{Max-Planck-Institut f\"ur Metallforschung, Heisenbergstr. 3,
D-70569 Stuttgart, Germany,} 
\affiliation{Institut f\"ur Theoretische und Angewandte Physik,
Universit\"at Stuttgart, Pfaffenwaldring 57, D-70569
Stuttgart, Germany}

\author{M. Rauscher}
\affiliation{Max-Planck-Institut f\"ur Metallforschung,
Heisenbergstr. 3, D-70569 Stuttgart, Germany,}
\affiliation{Institut f\"ur Theoretische und Angewandte Physik,
Universit\"at Stuttgart, Pfaffenwaldring 57, D-70569
Stuttgart, Germany}

\author{S. Dietrich}
\affiliation{Max-Planck-Institut f\"ur Metallforschung,
Heisenbergstr. 3, D-70569 Stuttgart, Germany,}
\affiliation{Institut f\"ur Theoretische und Angewandte Physik,
Universit\"at Stuttgart, Pfaffenwaldring 57, D-70569
Stuttgart, Germany}

\date{\today}

\begin{abstract}

We investigate the dynamics of nanoscale droplets in the vicinity
of chemical steps which separate parts of a substrate with different wettabilities. 
Due to long-ranged dispersion
forces, nanodroplets positioned on one side of the step perceive the different character of the 
other side even at some
distances from the step, leading to a dynamic response. The direction of the ensuing motion of such droplets does not only depend on the difference between the equilibrium
contact angles on these two parts but in particular on the difference between the corresponding Hamaker constants. Therefore the motion is not necessarily directed towards the more wettable side and can also be different from that of droplets which span the step.

\end{abstract}

\maketitle

The interest in dynamical wetting phenomena has increased
significantly with the development of micro- and nanofluidic
systems, which allow one to handle minute amounts of liquid containing,
e.g., DNA or proteins for chemical analysis and biotechnology
\cite{Karniadakis:2005, Dietrich:2005}. In particular chemically
heterogeneous systems with tailored, spatially varying wetting
properties have found important applications in this context
\cite{Daniel:1992,Kusumaatmaja:2007}. In open microfluidic systems,
fluids are transported on chemical channels, i.e., lyophilic
stripes embedded in lyophobic substrates.
While present devices are
based mostly on micron sized channels, further miniaturization down to the
nanoscale is clearly on the roadmap. This will eventually lead to
nanofluidic systems for which a variety of physical phenomena, 
which on the microscale and above are either irrelevant or summarized into boundary conditions, 
become important \cite{eijkel05,mukhopadhyay06}. 
For the optimization of the performance of nanofluidic systems it is 
critical to understand the basic fluidic issues occurring on those scales.
Recent theoretical studies of nanoscale fluids on chemically \cite{Cieplak:2006} and
topographically \cite{Moosavi:2006} structured substrates have
underscored the importance of such investigations.
 
Those analytical tools 
\cite{Greenspan:1978, Raphael:1988, Brochard:1989,Subramanian:2005} 
which rely on classical macroscopic theory are not adequate 
for this purpose. Apart from some molecular dynamics 
simulations \cite{Cieplak:2006,Yaneva:2004}, which are 
computationally demanding, to a large extent the available numerical 
investigations are based on solving thin film equations 
\cite{Schwartz:1998,Brusch:2002,Pismen:2006,Zhao:2006,Thiele:2006}. 
In most of theses studies the chemical heterogeneities are introduced  
via abrupt, lateral changes of those parameters which characterize the potentials of 
homogenous substrates 
\cite{Schwartz:1998,Pismen:2006,Zhao:2006}. However, this does not capture the actual
 behavior of such substrate potentials, even if the underlying chemical steps are taken to be atomically sharp \cite{Koch:1995}. 
Smooth chemical heterogeneities have been studied in Ref. 
\cite{Thiele:2006} by introducing a continuously varying Hamaker constant. 
But this approach is only applicable for very smooth variations of
the wetting properties. Studies of the dynamics of droplets in the
vicinity of topographic steps have shown that on the nanoscale a detailed modeling
of the substrate and thus of the resulting effective interface potential is mandatory \cite{Moosavi:2006}. For a chemical step, this has been worked out within the framework of
microscopic density functional theory with a view on the morphology of static wetting films \cite{Bauer:1999a,Bauer:1999b,Bauer:2000}.
\begin{figure}[b]
\centering
\includegraphics[width=0.9\linewidth,angle=0]{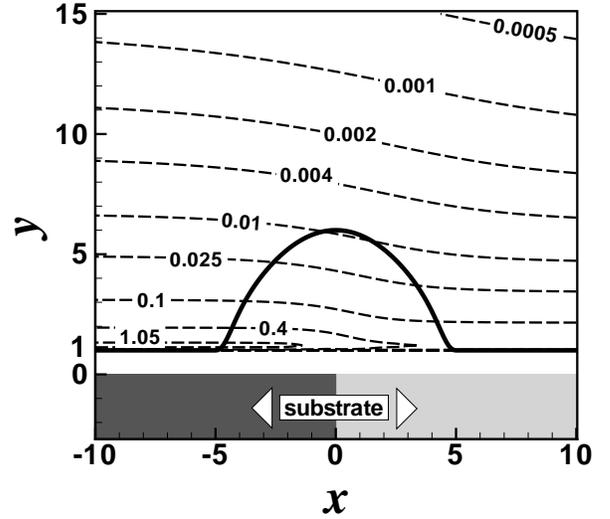}
\caption{\label{fig1}The chemical step is 
translationally invariant along the $z$ axis
(orthogonal to the image plane). A nanodroplet (full line) 
is exposed to the vertically and laterally varying DJP, the contour plot of which is shown.
 Here both sides of the substrate are taken to correspond to the $\ominus$ case with $b^{\lhd}=b^{\rhd}$ and $\theta_{eq}^{\lhd}=97.2^\circ$ 
($B^\lhd=0$, $C^\lhd=3$) and $\theta_{eq}^{\rhd}=51.3^\circ$ ($B^\rhd=0$, $C^\rhd=1$) [\,see Eqs.~(\ref{eq3}) and ~(\ref{eq4}) and the main text\,]. Lengths are measured in units  
of $b^{\lhd}$ (see the main text). A coating of the substrate is not indicated as here $B^{\lhd}=B^{\rhd}=0$ so that for $x\rightarrow\pm\infty$ the equilibrium thicknesses of the underlying film are the same, i.e., $y_0^\lhd=y_0^\rhd=y_0=b^{\lhd}$.}
\end{figure}

Here we study the dynamics of nanodroplets in the
vicinity of chemical steps, i.e., sharp and straight boundaries between two parts of a substrate with different wetting properties. Such
steps appear in open micro- and nanofluidic systems as the edges of
chemical channels. As a paradigmatic case 
we focus on a chemical step formed by two adjacent 
quarter spaces composed of different substrate particles. 
We analyze the driving force on the
droplets and perform numerical calculations employing the
standard boundary integral method for hydrodynamic Stokes flow providing 
the underlying dynamics. 

As depicted in Fig.~\ref{fig1} we consider a partially wetting, non-volatile, and incompressible
liquid forming a nanodroplet on top of a precursor wetting film over a chemical step. 
The type of chemical step which we consider here 
can be viewed as being composed of two different 
quarter spaces with each of the two corresponding upper surfaces coated additionally with a
 thin layer of different materials. Accordingly, as a basic element we 
first consider an edge without surface coating, say the left part of the substrate shown 
in Fig.~\ref{fig1} which we denote by the superscript $\lhd$. 
Assuming Lennard-Jones type intermolecular pair potentials $V_{\alpha
\beta}(r)={M_{\alpha \beta}}/{r^{12}}-{N_{\alpha
\beta}}/{r^6}$, where $M_{\alpha \beta}$
and $N_{\alpha \beta}$ are material parameters, 
and $\alpha$ and $\beta$ 
relate to liquid ($l$), substrate ($s$), or coating ($c$) particles,
the local disjoining pressure (DJP) corresponding to an $e$dge occupying
 $\Omega_s^{\lhd}=\lbrace {\bf{r}} \in \mathbb{R}^3
\mid x\leq 0,y\leq 0 , z\in \mathbb{R} \rbrace $ is given by \cite{Robbins:1991}\\
\begin{equation}
\label{eq1}
\Pi^{\triangleleft}_e(x,y)= \int_{\Omega^{\triangleleft}_s} 
\frac{\Delta M^{\lhd}}{\left|\vct{r}-\vct{r}'\right|^{12}}\,d^3r' -
\int_{\Omega^{\lhd}_s} 
\frac{\Delta N^{\lhd}}{\left|
\vct{r}-\vct{r}'\right|^6 }\,d^3r',
\end{equation} \\
\noindent
with $\Delta M^{\lhd}= \rho_l^2 M_{ll}-\rho_l\rho^{\lhd}_s M_{sl}^{\lhd}$ 
and $\Delta N^{\lhd}=\rho_l^2
N_{ll}-\rho_l\rho^{\lhd}_s N_{sl}^{\lhd}$\/ where $\rho_l$ 
and $\rho^{\lhd}_s$ are the
number densities of the liquid and the substrate, respectively. 
Because of its low density in Eq.~(\ref{eq1}) the effect of the vapor or gas phase has
been neglected.
The edge geometry allows one to analytically calculate 
both integrals in Eq.~(\ref{eq1}) which we denote 
as $\Pi_e^{12\lhd}$ and  $\Pi_e^{6\lhd}$, respectively, so that $\Pi_e^{\lhd}(x,y)=\Pi_e^{12\lhd}(x,y)-\Pi_e^{6\lhd}(x,y)$.  

Likewise, the contribution to the DJP of a thin 
$c$oating layer of thickness $d^{\lhd}$ on the $u$pper part of an edge $\Omega_c^{u\lhd}=\lbrace {\bf{r}} \in \mathbb{R}^3
\mid x\leq 0,-d^{\lhd}\leq y\leq0 , z\in \mathbb{R} \rbrace$ can be determined by assuming also a van der Waals 
type interaction between the coating and the liquid particles, i.e., $\Pi^{u\lhd}_c(x,y)=-\int_{\Omega_c^{u\lhd}} 
{\Delta N'^{\lhd}}/{\left|
\vct{r}-\vct{r}'\right|^6 }\,d^3r'$, with $\Delta N'^{\lhd}=\rho_l^2
N_{ll}-\rho_l\rho^{\lhd}_c N^{\lhd}_{cl}$\/ and $\Pi_{c}^{u\lhd}(x\rightarrow -\infty,y)=-\pi\Delta N'^{\lhd}d^{\lhd}/(2y^4)$ \cite{Moosavi:2006}. As a simplification we have neglected the effect of the repulsive 
part of the liquid-coating interaction which gives rise to a contribution  
shorter ranged ($\sim y^{-10}$) than the corresponding 
one  $\Pi_e^{12\lhd}\sim y^{-9}$ \cite{Bauer:1999a, Dietrich:1991}\/. The DJP of an edge including the coating of its upper side is 
$\Pi^{\lhd}_{ce}(x,y)=\Pi^{\lhd}_e(x,y+d^{\lhd})+\Pi^{u\lhd}_c(x,y)$.
For $x\rightarrow -\infty$ and $d^{\lhd}$ much smaller than the thickness of the wetting layer $y_0^{\lhd}$, the DJP of the coated edge 
reduces to that of a coated, laterally $h$omogeneous substrate:
$\Pi^{\lhd}_{ch}(y)={\pi\Delta M^{\lhd}}/{(45y^9)}-{\pi\Delta
N^{\lhd}}/{(6y^3)}+{\pi \Delta N''^{\lhd} d^{\lhd}}/{(2y^4)}$, with $\Delta N''^{\lhd}=  
\Delta N^{\lhd}- \Delta N'^{\lhd}=\rho_l(\rho_c^{\lhd}N_{cl}^{\lhd}-\rho_s^{\lhd} N^{\lhd}_{sl})$; $\Pi^{\lhd}_{ch}(y=y_0^{\lhd})=0$.

At this point we introduce dimensionless quantities (marked by a star) 
such that lengths are measured in units of 
$b^{\lhd}={[2\Delta M^{\lhd}/(15\,|\Delta N^{\lhd}|)]}^{1/6}$, which 
for $\Delta N^{\lhd}>0$ is the equilibrium wetting film thickness on the
$un$coated homogeneous substrate. 
The DJP is measured in units of $\sigma/b^{\lhd}$ 
where $\sigma$ is the liquid-vapor surface tension.
Thus the dimensionless DJP $\Pi_{ch}^{*\lhd}=\Pi_{ch}b^{\lhd}/\sigma$ 
far from the edge $(x\rightarrow -\infty)$ has the following form:
\begin{equation}
\label{eq2}
\Pi^{*\lhd}_{ch}(y_*)={C^{\lhd}}\left(\frac{1}{{y_*}^9} \pm
\frac{1}{{y_*}^3} + \frac{B^{\lhd}}{{y_*}^4}\right);
\end{equation} 
\noindent
in the following we drop the stars.

In Eq.~(\ref{eq2}) $B^{\lhd}=\pi \Delta
N''^{\lhd} d^{\lhd}/(2\,A^{\lhd}\,b^{\lhd 4})$ 
quantifies the contribution of the coating and 
 $C^{\lhd}=A^{\lhd}\,b^{\lhd}/\sigma$ where $A^{\lhd}=\pi{(\Delta M^{\lhd}/45)}^{-1/2}{(|\Delta N^{\lhd}|/6)}^{3/2}$ 
measures the effect of the competing intermolecular forces relative to the surface tension 
of the liquid-vapor interface. Since a more refined analysis of the DJP beyond Eq.~(\ref{eq1}) 
yields $B\neq0$ even in the absence of a coating 
layer \cite{Dietrich:1991,Dietrich:1988}, in the following we 
consider $B$ as an independent disposable parameter. 
In the second term on the right 
hand side of Eq.~(\ref{eq2}) the upper plus (lower minus) sign corresponds 
to $\Delta N^{\lhd}<0$ $(\Delta N^{\lhd}>0)$. 
In the following we shall refer to these 
cases as the plus $\oplus$ and the minus $\ominus$ cases, respectively.
The dimensionless form of the DJP (in units of $\sigma/b^{\lhd}$) for a single edge coated on the upper
side is given by 
\begin{eqnarray}
\label{eq3}
\Pi_{ce}^{\lhd}(x,y)&=& C^{\lhd}\bigg\{
\frac{45\,\Pi_e^{12\lhd}(x,y)}{\pi\,\Delta M^{\lhd}}
\pm\frac{6\,\Pi_e^{6\lhd}(x,y)}{\pi\,|\Delta N^{\lhd}|} \nonumber \\
&&+\frac{2\,B^{\lhd}\,[-\Pi_c^{u\lhd}(x,y)]}{\pi\,\Delta
N'^{\lhd}}\bigg\}.
\end{eqnarray} 
\noindent
Here $\Pi_e^{12\lhd}$, $\Pi_e^{6\lhd}$, and $\Pi_c^{u\lhd}$ are measured in units of $\sigma/b^{\lhd}$ whereas $\Delta M^{\lhd}$, $\Delta N^{\lhd}$, and $\Delta N'^{\lhd}$ are taken in units of $\sigma(b^{\lhd})^8$, $\sigma(b^{\lhd})^2$, and $\sigma(b^{\lhd})^2$, respectively; $x$ and $y$ are measured in units of $b^{\lhd}$ so that $C^{\lhd}$ and $B^{\lhd}$ are dimensionless. The contact angle $\theta^{\lhd}_{eq}$ of macroscopic droplets and those values of $C^{\lhd}$ and $B^{\lhd}$, which give rise 
to partial wetting, are related via
$\cos\theta_{eq}^{\lhd}=1+\int_{y_0^{\lhd}}^\infty \Pi_{ch}^{\lhd}(y)\,dy$ with $y_0^{\lhd}$ given 
implicitly by $\Pi_{ch}^{\lhd}(y_0^{\lhd})=0$ 
\cite{Dietrich:1988,Moosavi:2006}\/ so that $y_0^{\lhd}(B^{\lhd}=0)=b^{\lhd}$. Within these admissible ranges of values for $C^{\lhd}$ and $B^{\lhd}$ (see the insets of Fig. 2 in Ref.~\cite{Moosavi:2006}), the actual contact angle of nanodroplets defined, e.g., via their slope of the point of inflection or by spherical extrapolation of their top cap towards the substrate, may differ from that of macroscopic droplets depending on the size of 
the nanodroplets and details of the DJP. 
With $q=b^{\rhd}/b^{\lhd}$ the dimensionless form of the DJP for the right edge is given by
\begin{eqnarray}
\label{eq4}
\Pi_{ce}^{\rhd}(x,y)&=& C^{\rhd}\bigg\{
\frac{45q^9\,\Pi_e^{12\lhd}(-x,y)}{\pi\,\Delta M^{\lhd}}
\pm\frac{6q^3\,\Pi_e^{6\lhd}(-x,y)}{\pi\,|\Delta N^{\lhd}|} \nonumber \\
&&+\frac{2q^4\,B^{\rhd}\,[-\Pi_c^{u\lhd}(-x,y)]}{\pi\,\Delta
N'^{\lhd}}\bigg\}.
\end{eqnarray}

Due to the additivity of the interatomic potentials used here, the 
DJP of the chemical step can be obtained by 
superimposing the DJP of the constitutional 
parts, i.e., the two edges coated on the upper side. 
Thus the DJP of the whole substrate with a chemical step is given by
$\Pi(x,y)=\Pi^{\lhd}_{ce}(x,y)+\Pi^{\rhd}_{ce}(x,y)$. 
Figure~\ref{fig1} shows the contour lines of the DJP across a chemical step for a typical
example. 
\begin{figure}[t]
\center
\includegraphics[width=0.9\linewidth]{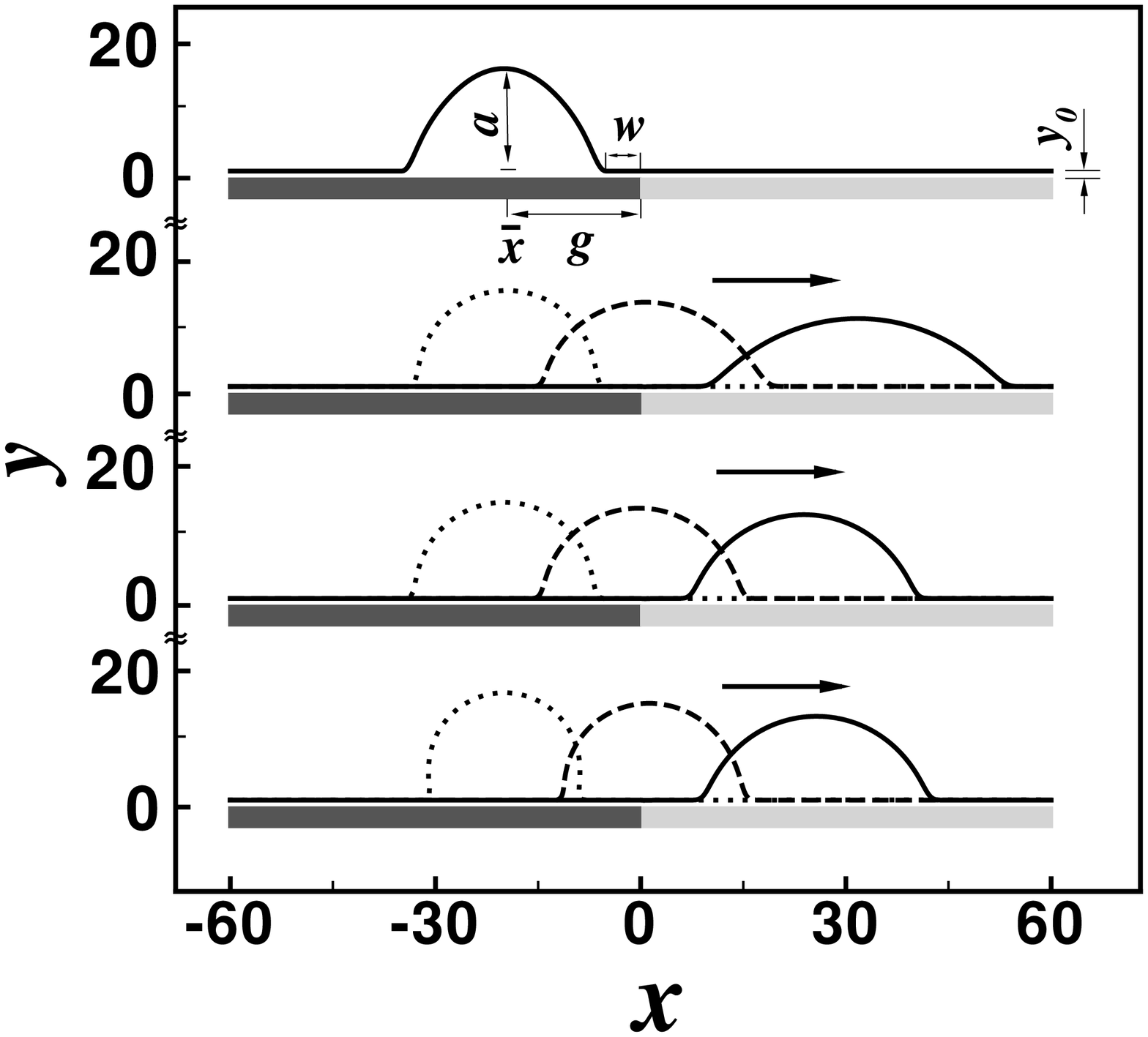}
\caption{The motion of a nanodroplet across three different chemical
steps for the
$\ominus$ case on both sides and for $B^\lhd=0$, $B^\rhd=0$, and $q=b^{\rhd}/b^{\lhd}=1.0$ so that $y_0^{\lhd}=y_0^{\rhd}=y_0$. 
The uppermost part shows the initial interface profile. The
parameters for the lower three parts are $[\:\theta_{eq}^{\lhd}=97.2^\circ (C^\lhd =3)$, $\theta_{eq}^{\rhd}=51.3^\circ (C^\rhd=1)]$, $[\:\theta_{eq}^{\lhd}=97.2^\circ (C^\lhd =3)$, $\theta_{eq}^{\rhd}=75.5^\circ (C^\rhd=2)]$, and $[\:\theta_{eq}^{\lhd}=120^\circ (C^\lhd =4)$, $\theta_{eq}^{\rhd}=75.5^\circ (C^\rhd=2)]$ from  top
to bottom. The profiles correspond to times $t=355$, 2000, and 22025; $t= 375$, 8125, and 24300;
and $t=185$, 7800, and 23500; respectively, in units of $\mu{b^{\lhd}}/(C^{\lhd}\sigma)$. $\lhd$ ($\rhd$) is the less (more) wettable substrate.}
\label{fig2}
\end{figure} 

In order to probe the influence of the DJP on droplets near and on
chemical steps we have performed mesoscopic 
hydrodynamic calculations based on the two-dimensional 
Stokes equation. In dimensionless form the continuity and Stokes equation
read $\bm{\nabla}\cdot\mathbf{u}=0\quad\text{and}\quad C^{\lhd}\:
\bm{\nabla}^2{\mathbf{u}}=\bm{\nabla}p$,
respectively, where $\mathbf{u}=(u_{x},u_{y})$ is the velocity
vector and  $p$ is the hydrostatic pressure. With the viscosity $\mu$, the velocity and time scales are taken to be $C^{\lhd}\,\sigma/\mu$ and
$\mu b^{\lhd}/(C^{\lhd}\sigma)$, respectively. Lengths and pressure are expressed in units of $b^{\lhd}$ and $\sigma/{b^{\lhd}}$, respectively. In the above Stokes equation the factor $C^{\lhd}$ appears because its dimensionless form has been obtained by rescaling with the parameters of the left hand side of the substrate. We have solved theses equations
numerically with a standard biharmonic boundary integral method
\cite{Kelmanson:1983}.
A no-slip boundary condition has been employed 
for the impermeable liquid-solid interface and it has been imposed that there is no flux through the
end sides of the system. 
Along the liquid-vapor interface it has been assumed that the tangential stresses 
are zero and that normal stresses are balanced by pressure, surface
tension, and the
disjoining pressure, i.e., ${\bf n}\cdot\tau\cdot{\bf n}=-p+\sigma\kappa+\Pi$ with the local curvature $\kappa$, the stress tensor $\tau$, and the unit surface normal vector ${\bf n}$.
The initial droplet shape has been taken to be a hemisphere which is smoothly
connected to the wetting layer, i.e.,
$y(x;t=0)= y_0^{\lhd}+a\lbrace 1-[(|x|-g)/a]^2]
\rbrace^{{\mid |x|-g\mid}^m+1}$ with
the droplet height $a$ in the center equal to half the base width and 
the distance $g$ of the droplet center from the boundary (see
Fig.~\ref{fig2}). In this study $m$ was chosen to be $10$\/. 
 
We first consider the case without coating ($B^\lhd=B^\rhd=0$, which implies that $\Delta N$ 
is negative on both sides corresponding to the $\ominus$ case).
Numerical results for different values of $C^{\rhd}$ and
$C^{\lhd}$ are shown in Fig.~\ref{fig2}. 
At time $t=0$ droplets of height $a=15$ have been positioned at a distance
$w=5$ (see Fig.~\ref{fig2}) from the  chemical step on the less wettable substrate. 
$C^{\lhd}$ was selected 
such that, during the \textit{initial} relaxation of the prepared droplet \textit{shape}, $w$ does not decrease. Thereafter, in all cases shown in Fig.~\ref{fig2}
the droplets 
gradually move towards the more wettable side ($\rhd$), as expected
intuitively, and continue their motion there.
This indicates that the nanodroplets 
can perceive the presence of the other part of the substrate over some lateral distances. As a function of time the wetting layer thickness changes slightly which is expected due to the Laplace pressure in the droplet and in view of the lateral boundary conditions on the flux.
\begin{figure}[t]
\centering
\includegraphics[width=0.49\linewidth]{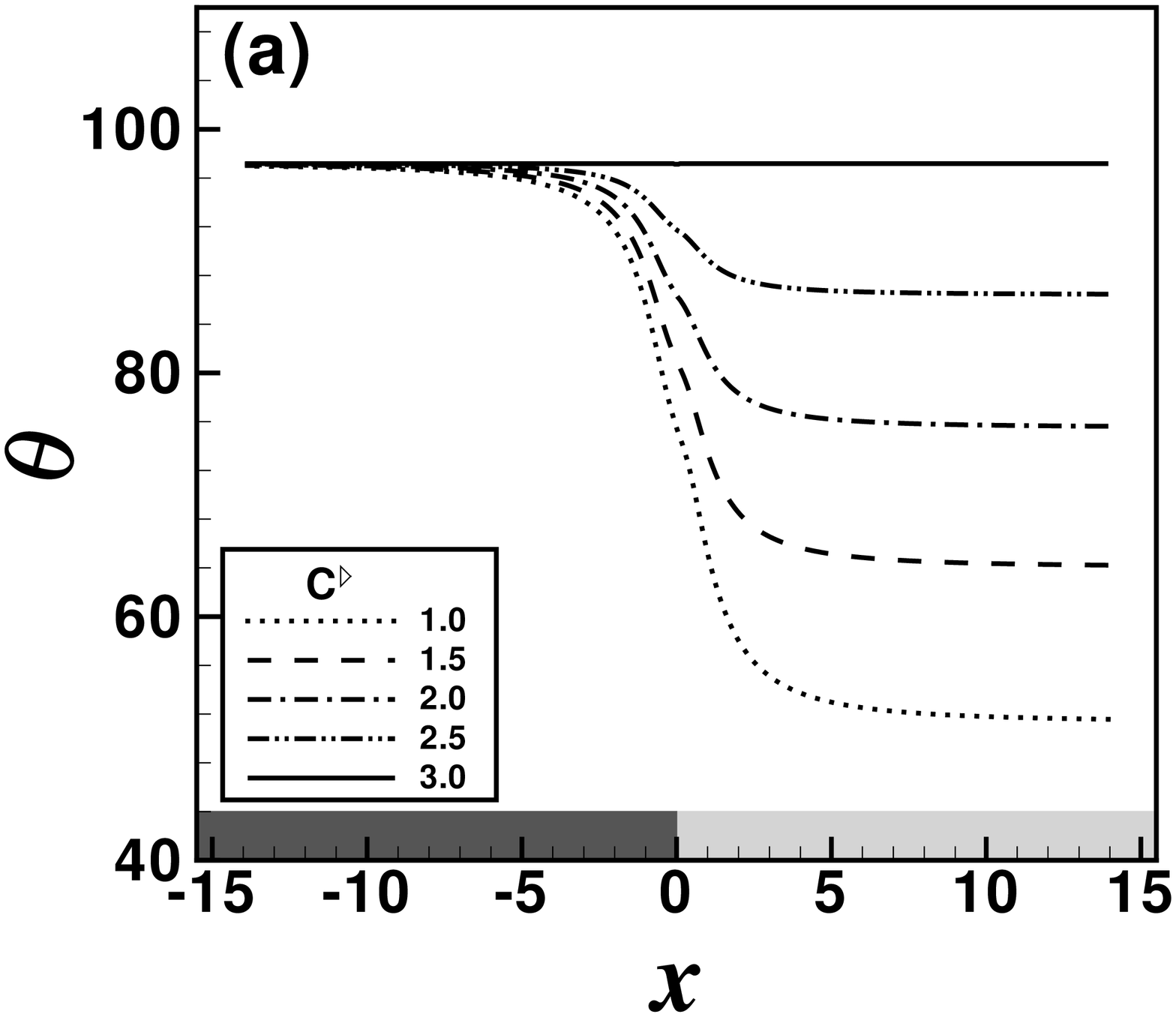}
\includegraphics[width=0.49\linewidth]{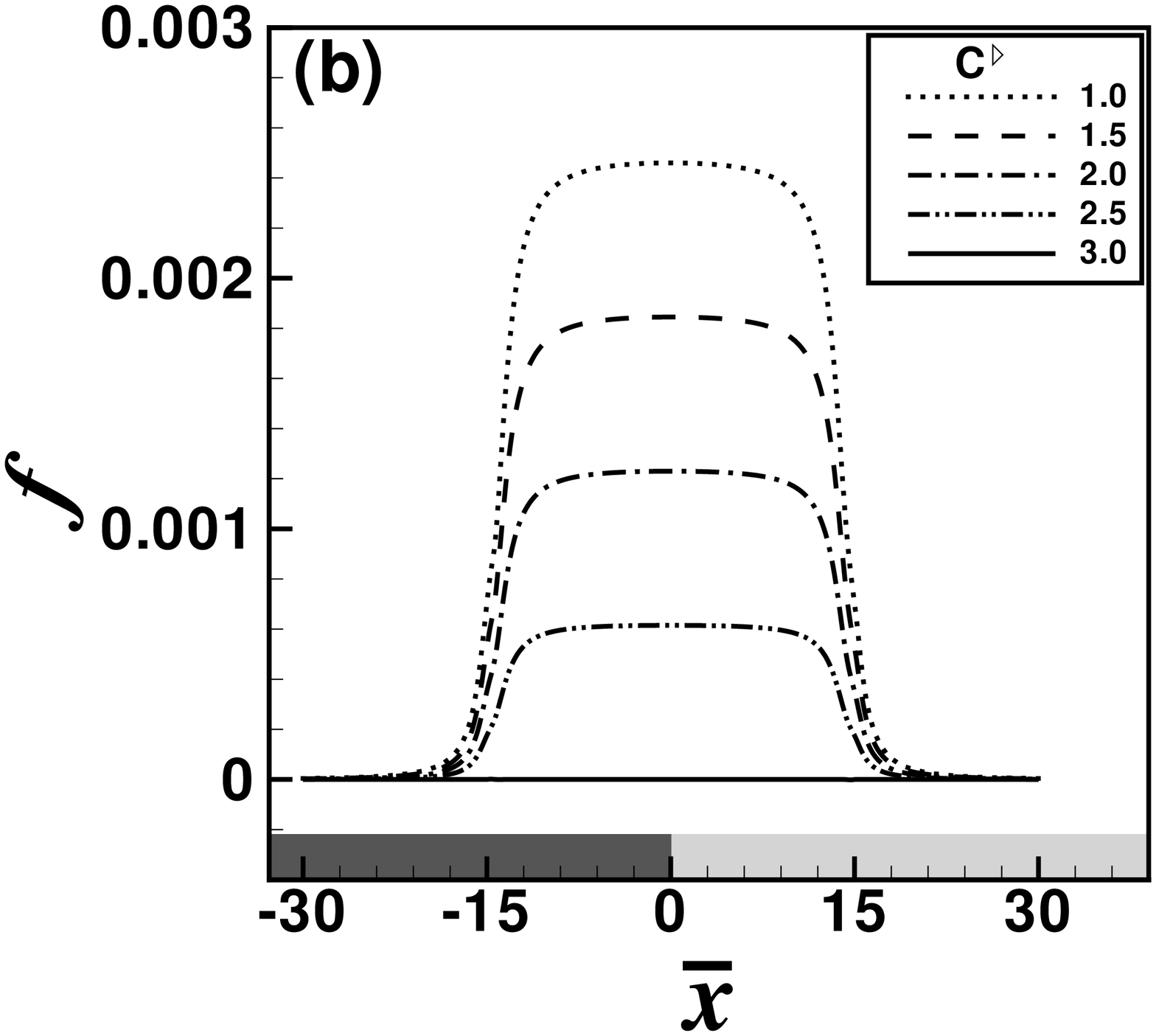}
\caption{\label{fig3}
(a) $\theta(x)$  and (b) dimensionless lateral force densities $f(\bar{x})$
for droplets with $a=15$ (for $\bar{x}$ see Fig.~\ref{fig2}) in the vicinity of a chemical step
separating two parts of the substrate with $C^\lhd=3$, $B^\lhd=B^\rhd=0$, and $q=1$\/. Both substrate sides correspond to the $\ominus$
case. $f$ is measured in units of $\sigma/{(b^{\lhd})}^2$.}
\end{figure}

On a homogeneous substrate, the equilibrium contact angle can be
calculated from the DJP. Extending this formula, one can define a
spatially varying ``contact angle'' via $\cos
\theta(x)=1+\int_{y_0(x)}^{\infty}\Pi(x,y)dy$, with
$\Pi(x,y_0(x))=0$. For $\Pi(x,y)$ rapidly varying as a function
of $x$ (e.g., in the close vicinity of a chemical step) and for very
small droplets, $\theta(x)$ is not the actual contact angle, but nonetheless,
as intuitively expected, $\partial_x\theta(x)$
is related to the lateral DJP induced force density (in units of $\sigma/(b^{\lhd})^2$) acting on a droplet: 
$f=1/{\Omega_d}\,
\times\int_{\partial\Omega_d}\Pi(x,y)\,n_{x}\,d{S}$ where
$\partial\Omega_d$ and $\Omega_d$ are the dimensionless droplet
surface and volume, respectively, and $n_{x}$ is the $x$-component 
of the outward surface normal ${\bf n}$. For the present purposes and for the sake of simplicity we have focused on nanodroplets large enough such that $\int_{y_0(x)}^a\Pi(x,y)dy\approx\int_{y_0(x)}^\infty \Pi(x,y)dy$. This means that for the types of the DJP considered in this study $a$ is taken to be at least 10 times the thickness of the underlying wetting film. Within this range, the phenomena discussed below are largely independent of the droplet size.
Since in the cases studied here the differences in wettability between the two parts of the
substrate are relatively small, in the following we have estimated $f(\bar{x})$ 
by considering droplets with the shape used as initial condition for the Stokes
dynamics but centered at $x=\bar{x}$\/.
For large and
symmetric droplets and if $\Pi(x,y)$
varies slowly over the width of the droplet such that
$\Pi(x_0+a,y)-\Pi(x_0-a,y)\approx 2\,a\,\partial_x \Pi(x_0,y)$,
$f$ and $-\partial \theta/\partial x$ have the same
sign. Thus droplets will move towards smaller $\theta(x)$, i.e., more wettable regions.
\begin{figure}[t]
\centering
\includegraphics[width=0.49\linewidth]{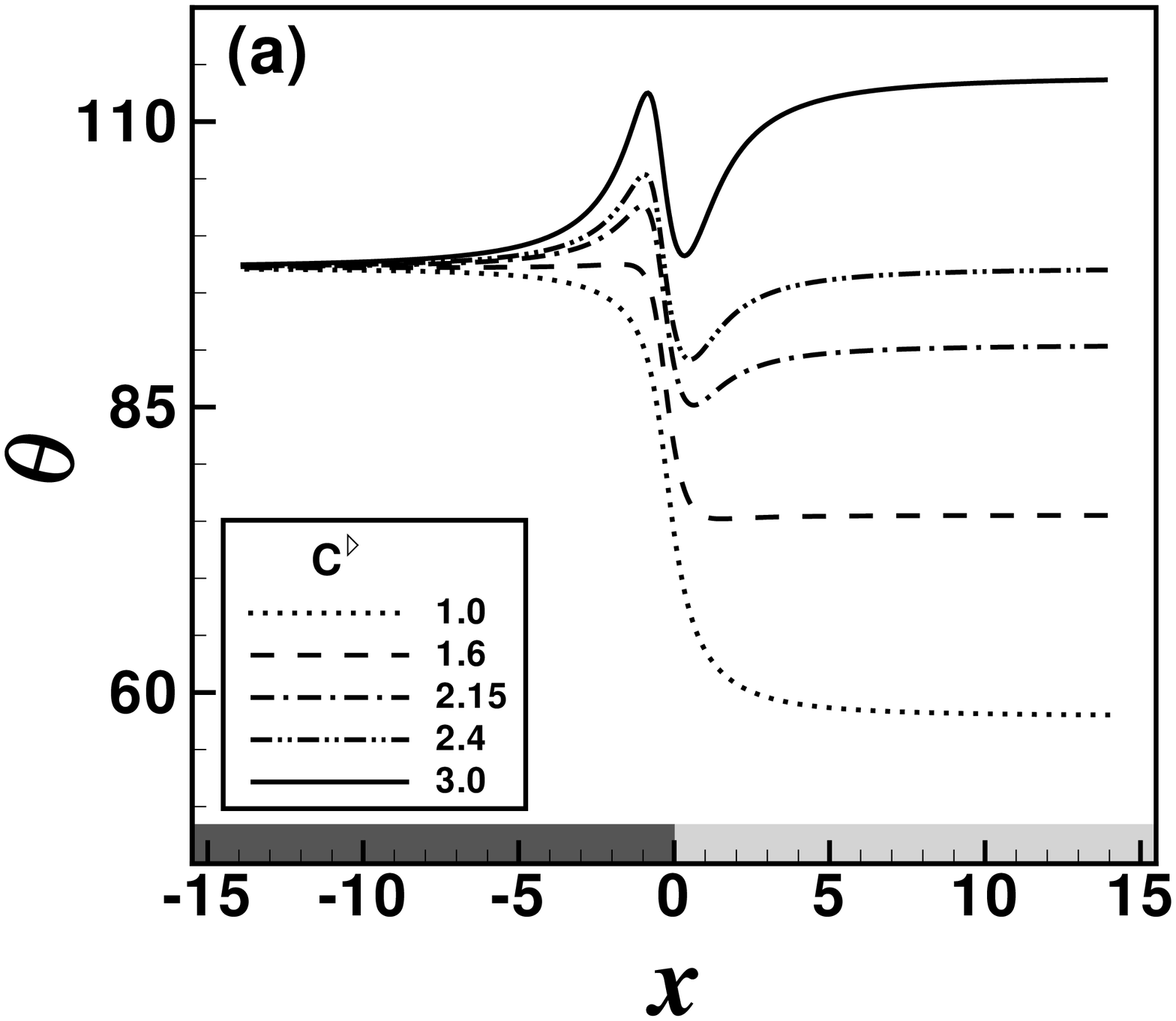}
\includegraphics[width=0.49\linewidth]{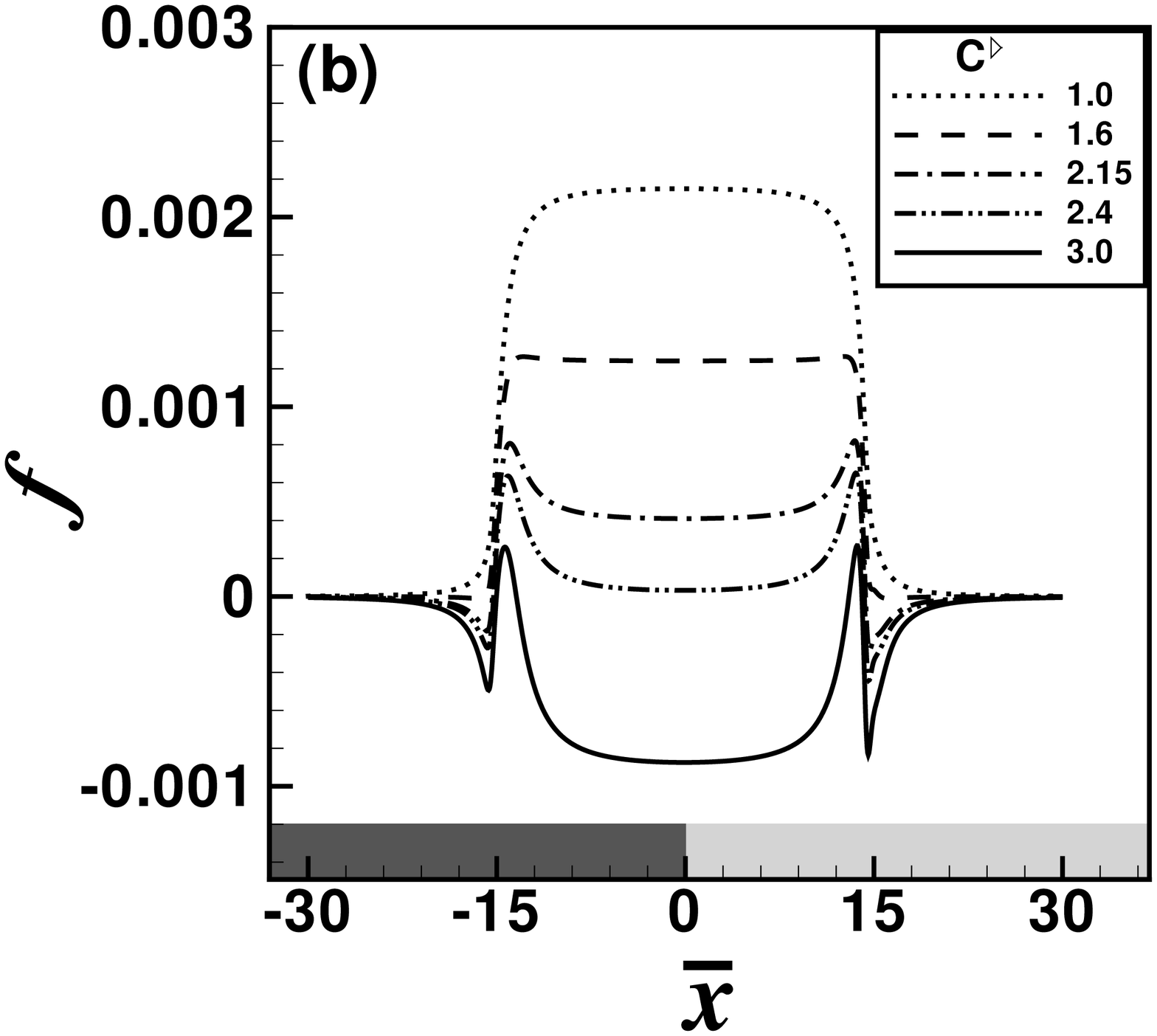}
\caption{\label{fig4} 
(a) $\theta(x)$ and (b) dimensionless lateral force densities $f(\bar{x})$ 
for droplets with $a=15$ in the vicinity of a chemical step
between two substrates with $C^\lhd=3$, $B^\lhd=B^\rhd=0$, and
$q=1.25$\/.  Both parts of the substrate correspond to the $\ominus$ case. $f$ is 
measured in units of $\sigma/{(b^{\lhd})}^2$.}
\end{figure}

Figure~\ref{fig3}(a) shows $\theta(x)$ in the vicinity of the
chemical step for the parameters used in Fig.~\ref{fig2}\/.
$\theta(x)$ monotonically decreases from $\theq^\lhd$ to $\theq^\rhd$
and, as shown in Fig.~\ref{fig3}(b), the force acting on droplets
of initial height $a=15$ is positive for all positions $\bar{x}$ of the
center of the drop. The force curve has a rather flat, plateau-like shape with a 
maximum at $\bar{x}=0$ and varies sharply if one of the
contact lines reaches the chemical step. This result is in agreement
with the numerical calculations and indicates that the droplets never
come to a complete stop.

In the previous example the local contact angle $\theta(x)$ changes
monotonically from $\theq^\lhd$ to $\theq^\rhd$\/. This is not
necessarily the case, in particular not 
for steps between materials generating different thicknesses of the wetting layer,
and for two substrates with different forms of the DJP, e.g., $\ominus$ on the
left and $\oplus$ on the right side. 
\begin{figure}[t]
\centering
\includegraphics[width=0.9\linewidth]{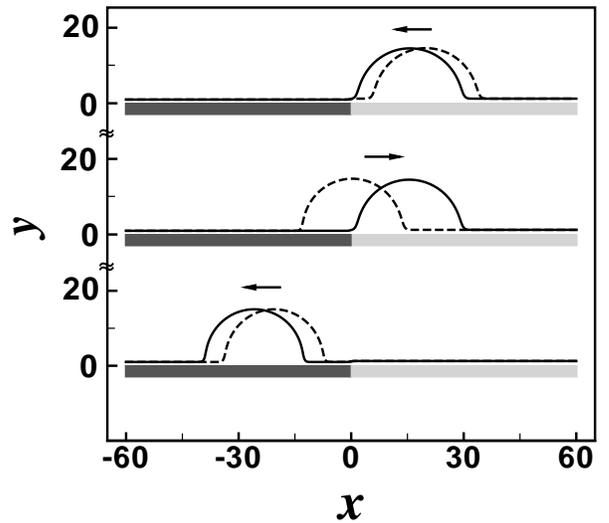}
\caption{\label{fig5} Motion of a nanodroplet near a 
chemical step in the $\ominus$ case on both sides, 
$B^\lhd=B^\rhd=0$, $C^\lhd=3$, $C^\rhd=2.15$, and $q=1.25$ 
(implying $\theta_{eq}^\lhd=97.2^\circ$ and $\theta_{eq}^\rhd=90.4^\circ$; compare Fig.~\ref{fig4}), 
starting at different positions.  
The droplet shapes are shown, top to bottom, at $t=170$, $100$, and $45$ (dashed)
and at $t=4750$, $4300$, and $40000$ (solid), respectively.
}
\end{figure}

Figure~\ref{fig4} shows $\theta(x)$  and $f(\bar{x})$ for $B^\lhd=B^\rhd =0$
and $q=1.25$\/. 
For $C^{\rhd}<1.5$, $\theta(x)$ decreases monotonically and $f(\bar{x})$ is 
positive for all $\bar{x}$. Therefore a droplet will move towards the
more wettable side. For $C^{\rhd}=1.6$ and larger, $\theta(x)$ is
nonmonotonic. 
The cases between $C^{\rhd}=1.6$ and $2.15$ are particularly interesting because, although $\theq^\lhd >
\theq^\rhd$, $\partial_x\theta >0$ outside a small region around the
step. A droplet is therefore expected to move towards the \textit{less}
wettable substrate. This is indeed the case because $f$ also changes
sign. Zeroes of $f(\bar{x})$ with $\partial_{\bar{x}}f(\bar{x})<0$ 
indicate that the chemical step can act as a pinning site for droplets.
For $C^{\rhd}>2.4$ the two parts of the substrate exchange roles in
that $\theq^\lhd < \theq^\rhd$. Almost everywhere $f(\bar{x})<0$ 
except for those values of $\bar{x}$ for which one of
the contact lines touches the step. Thus, a droplet is expected to move
towards the more wettable side. However, we expect pinning of the
contact lines at the step as the droplet moves from $\rhd$ to $\lhd$ because in this case 
$\partial_{\bar{x}}f(\bar{x})<0$  at the zeroes of $f(\bar{x})$.

In order to confirm these theoretical predictions we have performed a series of 
numerical calculations by positioning a 
nanodroplet at a distance $w=5$ on the right hand side  
of the chemical step. For the case $C^{\rhd}=2.15$, i.e., if the nanodroplet 
is initially positioned on the more wettable part of the substrate, the results of 
these calculations are shown in Fig.~\ref{fig5} 
for a nanodroplet with $a=15$. This droplet moves to 
the left (i.e., counterintuitively towards the less wettable substrate),
slows down, and stops when its advancing contact line touches the
chemical step.  A droplet initially spanning the step moves to
the right, as expected, but stops when its receding contact line
touching the step. Droplets completely located on the left part of the 
substrate move away from the chemical step, i.e., \textit{away} from the
\textit{more} wettable substrate, with the velocity decreasing as the
distance from the step increases. However, we do not observe a
complete stop. These numerical results are in complete agreement with our
analysis of the effective DJP induced force.
\begin{figure}[t]
\centering
\includegraphics[width=0.49\linewidth]{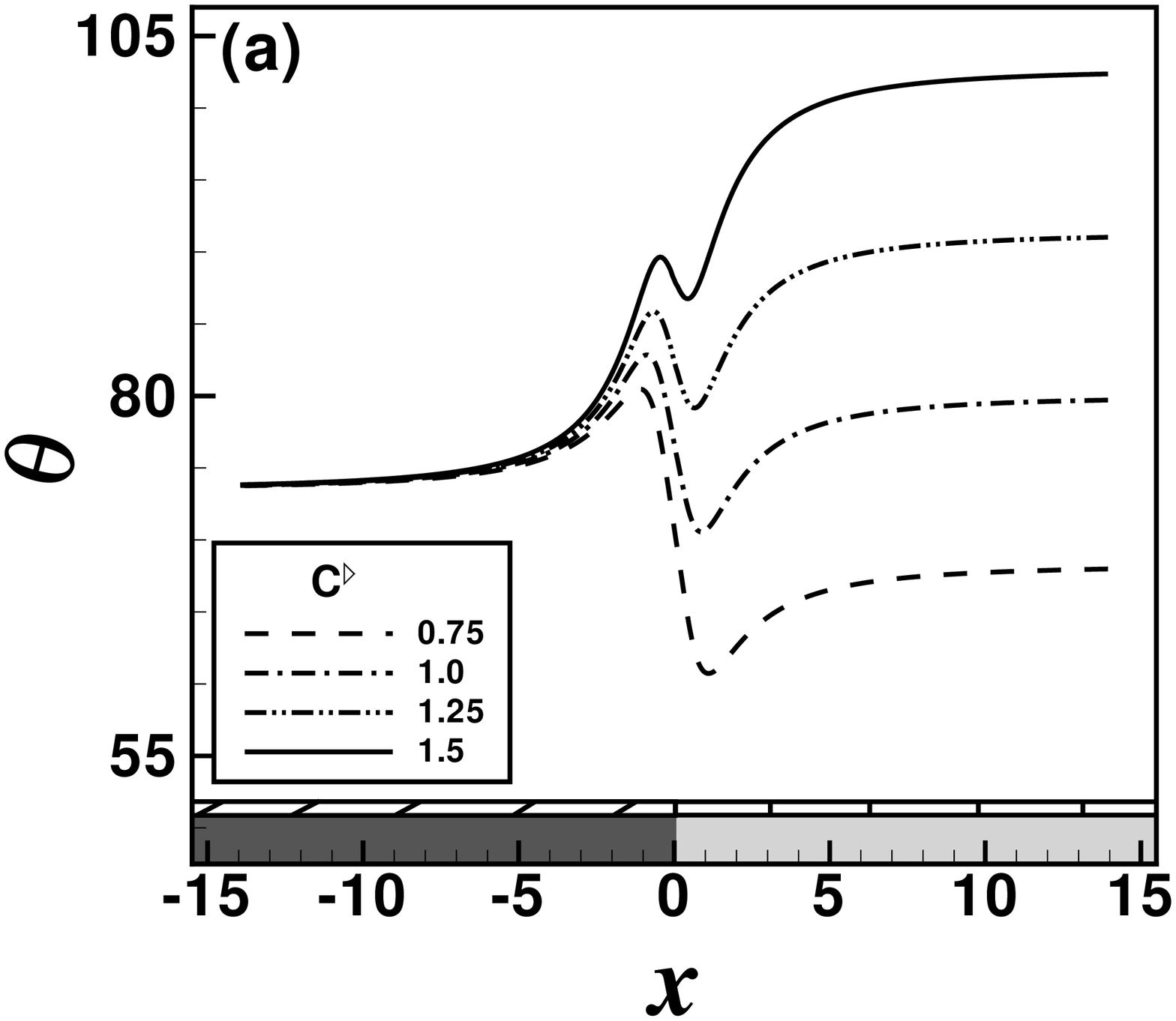}
\includegraphics[width=0.49\linewidth]{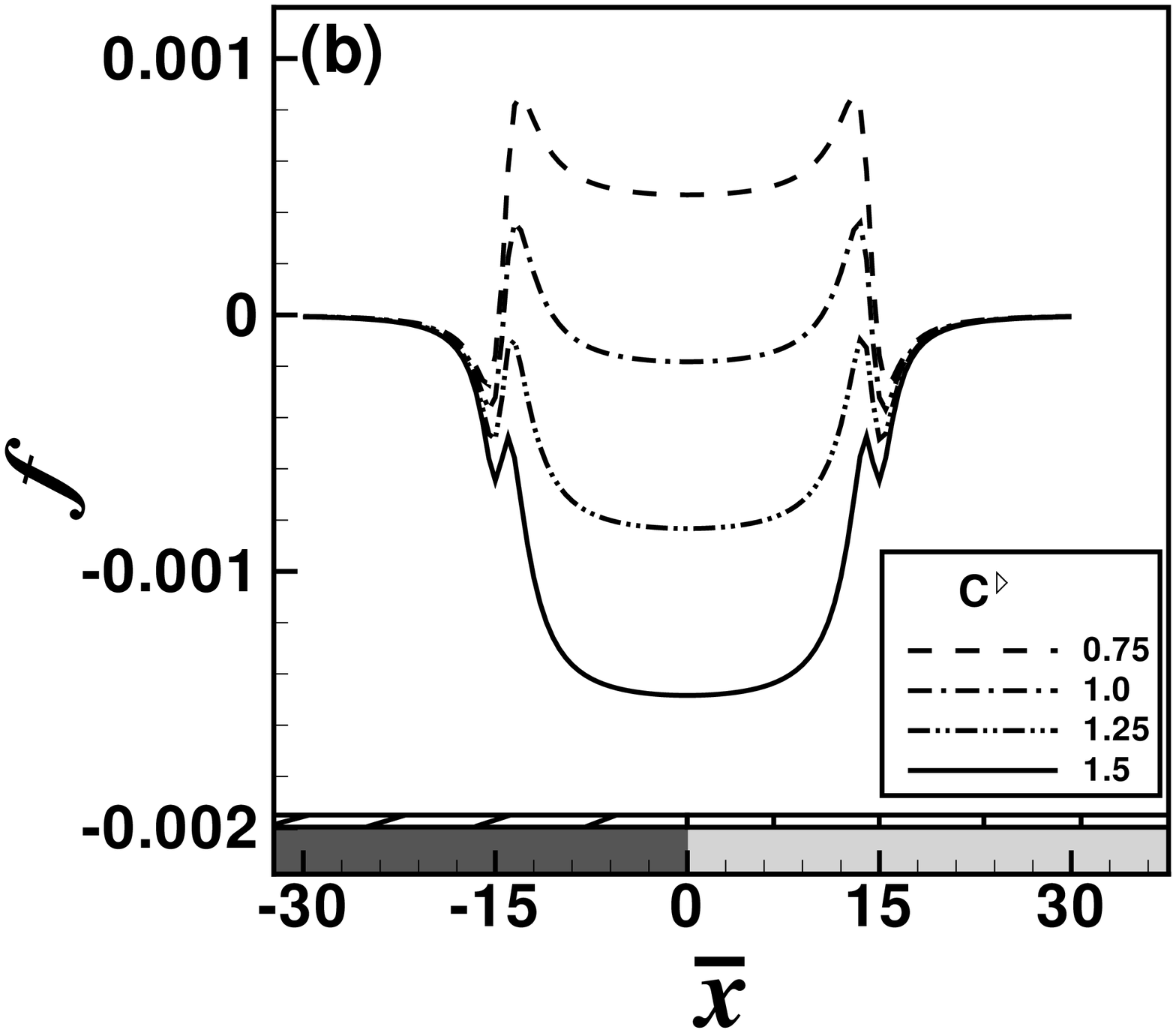}
\caption{\label{fig6} (a) $\theta(x)$ and (b) dimensionless lateral 
force densities $f(\bar{x})$ for droplets with $a=15$ in the vicinity of a
chemical step separating two substrates ($\oplus$ case on the left
and $\ominus$ case on the right) with $B^\lhd=-2.762$,
$C^\lhd=2$, $B^\rhd=-1$, and $q=1$ for different values of
$C^{\rhd}$. $f$ is measured in units of $\sigma/{(b^{\lhd})}^2$. The different coatings of the two parts of the substrate associated with the non-zero values of $B^{\lhd}$ and $B^{\rhd}$ are indicated in the figures. 
}
\end{figure}
Figure~\ref{fig6} shows an example for $q=1$ in which the droplets can
move towards the less wettable substrate. This is the case for
different signs of the long-ranged term (here $\oplus$ case on the
left and $\ominus$ case on the right) and for non-zero values of $B$
(here $B^\lhd=-2.762$, $C^\lhd=2$, and $B^\rhd=-1$). These values
are chosen such that the equilibrium wetting film thicknesses are equal on both sides of the substrate.
Outside a narrow region around the chemical step $\partial_x\theta(x)$ is positive everywhere. This indicates that the droplet will move
to the left independent of the relative values of $\theq^\lhd$ and $\theq^\rhd$, although 
it will be pinned at the step if coming from the right. This is
confirmed by the analysis of $f(\bar{x})$ in Fig.~\ref{fig6}(b). If
both contact lines are on the same side of the step,
$f(|\bar{x}|>a)<0$. If the droplets span the step, i.e., $|\bar{x}|<a$, the direction of
the motion is towards the more wettable part of the substrate.

In all the cases we considered the direction of motion of a droplet
near a chemical step (i.e., with both contact lines on the same
side of the step) is determined by $\partial_x \theta(x)$. The sign of
$\partial_x \theta(x)$ is the same as the sign of
$-\partial_x\Pi(x,y_0(x))$\/. Expanding $\Pi(x,y)$ for large $|x|$ up to
$O(|x|^{-3})$ one finds on both sides of the step the limiting value for a homogeneous
substrate.
This
means that the wetting film thickness $y_0$ is independent of
$x$ up to this order. To leading order the gradient of the equilibrium contact angle
with respect to $x$ is  the same on both sides of
the chemical step and given by
\begin{equation}
\label{eq5}
\partial_x\theta(x) \sim \frac{3}{4\,|x|^3}\,\left[-(\pm
q^3)\,C^\rhd \pm C^\lhd \right] + \mathcal{O}(|x|^{-4}),
\end{equation}
with the plus and minus signs in front of $C^\rhd$ and
$C^\lhd$ corresponding to the $\oplus$ and the $\ominus$ case on
the right and the left hand side, respectively. This means that asymptotically far from the step the droplets move in the same direction on both sides of the step.
For instance, for the case discussed in Fig.~\ref{fig4} (with $B^{\lhd}=B^{\rhd}=0$) the 
equilibrium contact angle can be shown to be larger on the right side of the step if
$q\,C^\rhd > C^\lhd$, i.e., for
$C^\rhd >2.4$ and the droplets move to the left for
$C^\rhd >C^\lhd/q^3 = 1.536$ and $|\bar{x}|>a$, in agreement with
the observations. 

In summary, we have outlined an approach which allows one to study in detail the behavior of 
nanodroplets near as well as on chemical heterogeneties. Our investigation reveals 
the dynamics of nanodroplets in the vicinity of chemical heterogenities 
caused by long-ranged forces.
We have shown that the direction of motion of
the droplets is, to leading order in the distance from the step,
determined by the competition of the van der Waals forces acting
between the droplet and the two different materials of the substrate, i.e., 
the difference in the Hamaker constants (see Eq. (\ref{eq5})), rather than the
equilibrium contact angles which depend also on the short-ranged parts of 
interaction potentials and on 
the subleading terms in the disjoining pressure.
If the van der
Waals forces direct the droplet towards the less wettable material,
the droplet will stop as soon as the advancing contact line hits
the step. Otherwise it will continue at a velocity rapidly
decreasing with the distance from the step.  Droplets which
span the chemical step will move towards the more wettable
substrate; however, the receding contact line can be pinned by the
step.

This study demonstrates that taking into account the effect of 
long-ranged intermolecular forces is mandatory for accurately controlling
and guiding the liquids in open nanofluidic systems. 
Recent experiments have shown that the arrangement of droplets
on structured substrates can be explained by their crossing of chemical steps from
the less wettable to the more wettable side \cite{Leopoldes:2005}.
Our study indicates that in general there can be free-energetic barriers to this process which would result in
significantly altered patterns. Our analysis provides a microscopic
approach to the pinning and depinning of three-phase contact lines 
at chemical surface heterogeneities which goes beyond the macroscopic
picture of a sharp transition between regions of different
wettability on a substrate
\cite{casagrande89,dossantos96,ondarcuhu95,raphael89,shanahan99}
or the phenomenological mesoscopic approach of introducing lateral variations of the
parameters entering into the effective interface potential \cite{thiele06}.

{\bf Acknowledgment.} M. R. acknowledges financial support by the
Deutsche Forschungsgemeinschaft (DFG) within the priority
program SPP 1164 under grant number RA {1061/2-1}.

\end{document}